# Three-Receiver Broadcast Channels with Side Information

Saeed Hajizadeh and Ghosheh Abed Hodtani

*Abstract*—Three-Receiver broadcast channels (BC) are of interest due to their information-theoretic differences with two-receiver one. In this paper, we derive achievable rate regions for two classes of 3-receiver BC with side information (SI), i.e. Multilevel BC (MBC) and 3-receiver less noisy BC, using a combination of superposition coding, Gelf'and-Pinsker binning scheme and Nair-El Gamal indirect decoding. Our rate region for MBC subsumes Steinberg's rate region for 2-receiver degraded BC with SI as its special case. We will also show that the obtained achievable rate regions in the first two cases are tight for several classes of non-deterministic, semi-deterministic, and deterministic 3-receiver BC when SI is available both at the transmitter and at the receivers. We also prove that as far as a receiver is deterministic in the three-receiver less noisy BC, the presence of side information at that receiver does not affect the capacity region. We have also provided the writing on dirty paper (WDP) property for 3-receiver BC is provided as an example. In the last section, we provide simple bounds on the capacity region of the Additive Exponential noise three-receiver broadcast channels with Additive Exponential interference (AEN-3BC-EI).

## I. INTRODUCTION

BCs are one of the most important channels in multiuser information theory and have been broadly studied since being introduced by Cover [1] in 1972. Bergmans used superposition coding to find an achievable rate region for degraded two-receiver BC [2] which was proved to be optimal by Gallager [3] and Ahlswede and Körner [4]. The capacity region of special classes of BC has been obtained until now and can be found in [5], [6], Körner and Marton [7], and [8]. The best known inner and outer bound on the capacity region of general two-receiver BC, however, is given by Marton [9] and Nair-El Gamal [10], respectively. Unlike multiple access channels (MACs), extension of the results for two-receiver BC to BCs with more than two receivers is not in general optimal.

Saeed Hajizadeh is with the Electrical Engineering Department, Ferdowsi University of Mashhad, Mashhad, Iran. (e-mail: saeed.hajizadeh1367@gmail.com).

Ghosheh Abed Hodtani, is with the Electrical Engineering Department, Ferdowsi University of Mashhhad, Mashhad, Iran. (e-mail: ghodtani@gmail.com).

The k-receiver, $k \geq 3$, BC was first studied by Borade *et al.* in [11] where they simply surmised that straightforward extension of Körner-Marton's capacity region for two-receiver BCs with degraded message sets [7] to k-receiver multilevel broadcast networks is optimal. Nair-El Gamal [12] showed that the capacity region of a special class of 3-receiver BCs with two degraded message sets when one of the receivers is a degraded version of the other, is a superset of [11], thus proving that direct extension of [7] is not in general optimal. Nair and Wang later in [13] established the capacity region of the 3-receiver less noisy BC.

Channels with SI, were first studied by Shannon [14], where he found the capacity region of the Single-Input-Single-Output channel when SI is causally available at the encoder. Gelf'and and Pinsker [15] found the capacity of a single-user channel when SI is non-causally available at the transmitter while the receiver is kept ignorant of it. Cover and Chiang [16] extended the results of [15] to the case where SI is available at both the encoder and the decoder. Multiple user channels with SI were studied in [17] where inner and outer bounds on the capacity region of degraded BC with non-causal SI and capacity region of degraded BC with causal SI were found. Moreover, [18], added SI to [9]. Khosravi and Marvasti [19] added SI, both to [9] and [10] and their result contains that of [18]. Lapidoth and Wang [20] found the capacity of semi-deterministic two receiver broadcast channel with SI.

Gaussian channels with SI were studied by Max H. Costa [21]. He proved that a Gaussian single-user channel capacity is not afflicted by an extra additive Gaussian i.i.d interference as long as we have full knowledge of the extra interference at the encoder so that we can optimize our transmitter to achieve the Gaussian channel capacity with no interference. Young-Han Kim *et al.* extended [21] to degraded BC, MAC, and Relay Channel (RC) [22]. Reza Khosravi-Farsani [23] also extended [21] to the two-way channel, which was first studied by Shannon [24].

Exponential noise was first studied by Anantharam and Verdú [31] where they characterized the capacity of the Additive Exponential Noise (AEN) point-to-point channel. Verdú [32] also characterized the capacity region of the AEN multiple access channel with independent sources. Hajizadeh and Hodtani [33] studied AEN-BC. Monemizadeh, Hajizadeh and Hodtani [34] found capacity bounds for the exponentially dirty paper.

In this paper, we first find the achievable rate region of MBC and 3-receiver less noisy BC both with SI non-causally available at the encoder. Our achievable rate regions reduce to that of [12] and [13] when there is no SI. Our achievable rate region for MBC also subsumes Steinberg's achievable rate region for 2-receiver degraded BC with SI as its special case. The achievable rate regions are tight for several classes of non-deterministic, semi-deterministic and deterministic MBC and 3-receiver less noisy BC when SI is available both at the transmitter and at all the receivers. WDP property for three-receiver BCs is provided as an example. We then find an upper bound on the capacity region of AEN-3BC-EI.

The rest of the paper is organized as follows. In section II, basic definitions and preliminaries are presented. In section III, MBC with SI is studied while section IV is devoted to 3-receiver less noisy BCs with SI. In section V, examples are given. The AEN-3BC-EI is studied in section VI where in section VII, conclusion is given.

## II. BASIC DEFINITIONS AND PRELIMINARIES

### A. Basic definitions

Random variables and their realizations are denoted by uppercase and lowercase letters, respectively, e.g. $x$ is a realization of $X$. Let $\mathcal{X}, \mathcal{Y}_1, \mathcal{Y}_2, \mathcal{Y}_3$ and $\mathcal{S}$ be finite sets showing alphabets of random variables. The n-sequence of a random variable is given by $X^n$ where the superscript is omitted when the choice of $n$ is clear, thus we only use boldface letters for the random variable itself, i.e. $\boldsymbol{x} = x^n$. Throughout, we assume that $X_i^n$ is the sequence $(X_i, X_{i+1}, \ldots, X_n)$.

***Definition 1:*** A channel $X \to Z$ is said to be a degraded version of the channel $X \to Y$ with SI if $X \to Y \to Z$ is a Markov chain conditioned on every $s \in \mathcal{S}$ with $p(y, z|x, s)$ having the same marginal distributions.

MBC with SI, denoted by $(\mathcal{X}, \mathcal{S}, \mathcal{Y}_1, \mathcal{Y}_2, \mathcal{Y}_3, p(y_1, y_3|x, s), p(y_2|y_1))$, is a 3-receiver BC with 2-degraded message sets with input alphabet $\mathcal{X}$ and output alphabets $\mathcal{Y}_1, \mathcal{Y}_2$, and $\mathcal{Y}_3$. The SI is the random variable $S$ distributed over the set $\mathcal{S}$ according to $p(s)$. The transition probability function $p(y_1, y_3|x, s)$ describes the relationship between channel input $X$, side information $S$, and channel outputs $Y_1$ and $Y_3$ while the probability function $p(y_2|y_1)$ shows the virtual channel modeling the output $Y_2$ as the degraded version of $Y_1$. Independent message sets $m_0 \in \mathcal{M}_0$ and $m_1 \in \mathcal{M}_1$ are to be reliably sent, $m_0$ being the common message for all the receivers and $m_1$ the private message only for $Y_1$. Channel model is depicted in Fig. 1.

***Definition 2:*** A $(n, 2^{nR_0}, 2^{nR_1}, \epsilon)$ two-degraded message set code for MBC with SI $(p(y_1, y_3|x, s), p(y_2|y_1))$ consists of an encoder map

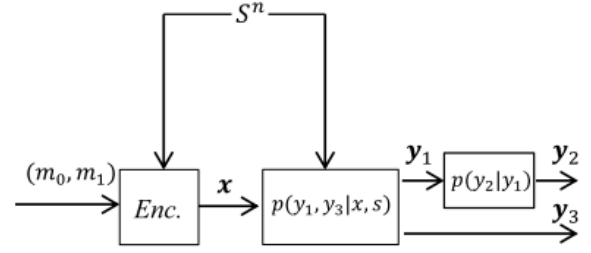

Figure 1. Multilevel broadcast channel with side information.

$$f : \{1,2, \ldots, M_0\} \times \{1,2, \ldots, M_1\} \times \mathcal{S}^n \to \mathcal{X}^n$$

and a tuple of decoding maps

$$g_{y_1} : \mathcal{Y}_1^n \to \{1,2, \ldots, M_0\} \times \{1,2, \ldots, M_1\}$$
$$g_{y_2} : \mathcal{Y}_2^n \to \{1,2, \ldots, M_0\}$$
$$g_{y_3} : \mathcal{Y}_3^n \to \{1,2, \ldots, M_0\}$$

such that $P_e^{(n)} \leq \epsilon$, i.e.

$$\frac{1}{M_0 M_1} \sum_{m_0=1}^{M_0} \sum_{m_1=1}^{M_1} \sum_{s^n \in \mathcal{S}^n} p(s) p\{g_{y_1}(\boldsymbol{y}_1) \neq (m_0, m_1) \text{ or }$$
$$g_{y_2}(\boldsymbol{y}_2) \neq m_0 \text{ or } g_{y_3}(\boldsymbol{y}_3) \neq m_0 | \boldsymbol{s}, \boldsymbol{x}(m_0, m_1, \boldsymbol{s})\} \leq \epsilon$$

The rate pair of the code is defined as

$$(R_0, R_1) = \frac{1}{n}(\log M_0, \log M_1)$$

A rate pair $(R_0, R_1)$ is said to be $\epsilon$-achievable if for any $\eta > 0$ there is an integer $n_0$ such that for all $n \geq n_0$ we have a $(n, 2^{n(R_0-\eta)}, 2^{n(R_1-\eta)}, \epsilon)$ code for $(p(y_1, y_3|x, s), p(y_2|y_1))$.

The union of the closure of all $\epsilon$-achievable rate pairs is called the capacity region $\mathcal{C}_{MBC}$.

***Definition 3:*** A channel $X \to Y$ is said to be less noisy than the channel $X \to Z$ in the presence of SI if

$$I(U; Y|S = s) \geq I(U; Z|S = s)$$
$$\forall p(u, x, y, z|s) = p(u|s) p(x|u, s) p(y, z|x, s) \text{ and } \forall s \in \mathcal{S}.$$

The 3-receiver less noisy BC with SI is depicted in Fig. 2, where $Y_1$ is less noisy than $Y_2$ and $Y_2$ is less noisy than $Y_3$, i.e. according to [13], $Y_1 \succcurlyeq Y_2 \succcurlyeq Y_3$.

The messages $m_1 \in \mathcal{M}_1$, $m_2 \in \mathcal{M}_2$, $m_3 \in \mathcal{M}_3$ are to be reliably sent to receivers $Y_1, Y_2, \text{ and } Y_3$, respectively. The code and rate tuple definitions are as follows

$$(n, 2^{nR_1}, 2^{nR_2}, 2^{nR_3}, \epsilon)$$
$$(R_1, R_2, R_3) = \frac{1}{n}(\log M_1, \log M_2, \log M_3)$$

Achievable rate tuples and the achievable rate region and the capacity region $\mathcal{C}_L$ are defined in just the same way as MBC with SI.

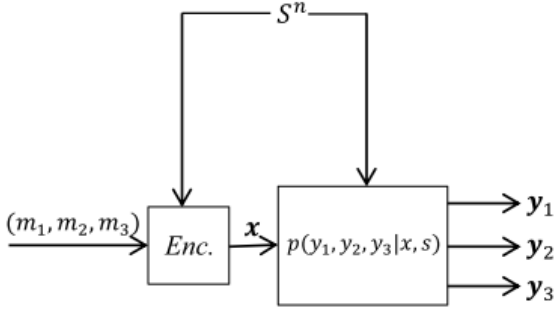

Figure 2. Three-receiver less noisy broadcast channel with side information.

## B. Preliminaries

Consider a sequence of independent and identically distributed random variables $X = (X_1, X_2, ..., X_n)$ each distributed according to $p(x)$. A sequence $x = (x_1, x_2, ..., x_n)$ is said to be $\epsilon$-typical if

$$\left|-\frac{1}{n}\log p(x) - H(X)\right| \leq \epsilon$$

Now let us define the typical set to be the set of all $\epsilon$-typical n-sequences $x$.

By the law of large numbers (LLN), [25], we have

$$-\frac{1}{n}\log p(x) = -\frac{1}{n}\log\prod_{i=1}^{n} p(x_i) \to H(X), \text{in probability}$$

Therefore we have

1) $p(x \in A_\epsilon^{(n)}) \to 1$ as n tends to infinity
2) $2^{-n(H(X)+\epsilon)} \leq p(x \in A_\epsilon^{(n)}) \leq 2^{-n(H(X)-\epsilon)}$
3) $(1-\epsilon)2^{n(H(X)-\epsilon)} \leq \| A_\epsilon^{(n)} \| \leq 2^{n(H(X)+\epsilon)}$

*Proof:* A proof can be found in [26].

Now consider a random variable $Y$ that is jointly typical with $X$ according to some $p(x,y)$. We say that $(x,y) \in A_\epsilon^{(n)}(X,Y)$, i.e. $(x,y)$ are jointly $\epsilon$-typical if

$$\left|-\frac{1}{n}\log p(x) - H(X)\right| \leq \epsilon$$
$$\left|-\frac{1}{n}\log p(y) - H(Y)\right| \leq \epsilon$$
$$\left|-\frac{1}{n}\log p(x,y) - H(X,Y)\right| \leq \epsilon$$

and by the LLN we have

1) $p((x,y) \in A_\epsilon^{(n)}) \to 1$ as n tends to infinity
2) $2^{-n(H(X,Y)+\epsilon)} \leq p((x,y) \in A_\epsilon^{(n)}) \leq 2^{-n(H(X,Y)-\epsilon)}$
3) $(1-\epsilon)2^{n(H(X,Y)-\epsilon)} \leq \| A_\epsilon^{(n)} \| \leq 2^{n(H(X,Y)+\epsilon)}$
4) $2^{-n(H(Y|X)+2\epsilon)} \leq p\left((y|x) \in A_\epsilon^{(n)}\right) \leq 2^{-n(H(Y|X)-2\epsilon)}$

*Proof:* A proof can be found in [26].

The concept of typicality can be extended to arbitrary number of random variables and an extension is provided in [27].

## III. MULTILEVEL BROADCAST CHANNEL WITH SIDE INFORMATION

Define $\mathcal{P}$ as the collection of all random variables $(U, V, S, X, Y_1, Y_2, Y_3)$ with finite alphabets such that

$$p(u,v,s,x,y_1,y_2,y_3) = p(s)p(u|s)p(v|u,s)p(x|v,s)p(y_1,y_3|x,s)p(y_2|y_1) \quad (1)$$

By (1), the following Markov chains hold

$$(U,V) \to (X,S) \to (Y_1, Y_3) \quad (2)$$
$$(S,X,Y_3) \to Y_1 \to Y_2 \quad (3)$$
$$U \to (V,S) \to X \quad (4)$$

*Theorem 1:* A pair of nonnegative numbers $(R_0, R_1)$ is achievable for MBC with SI non-causally available at the transmitter provided that

$$\begin{aligned} R_0 &\leq \min\{I(U;Y_2) - I(U;S), I(V;Y_3) - I(UV;S)\} \\ R_1 &\leq I(X;Y_1|U) - I(V;S|U) - I(X;S|V) \\ R_0 + R_1 &\leq I(V;Y_3) + I(X;Y_1|V) - I(X;S|V) - I(UV;S) \end{aligned} \quad (5)$$

for some $(U,V,S,X,Y_1,Y_2,Y_3) \in \mathcal{P}$.

*Corollary 1.1:* By setting $S \equiv \emptyset$ in (5), the achievable rate region in Theorem 1 is reduced to the achievable rate region of MBC given in [12].

*Corollary 1.2:* By setting $Y_3 = Y_1$ and $V = U$ in (5), our achievable rate region reduces to that of [17] for the two-user degraded BC with SI.

*Proof:* Fix n and a joint distribution on $\mathcal{P}$. Note that side information is distributed i.i.d according to

$$p(s) = \prod_{i=1}^{n} p(s_i)$$

Split the $\mathcal{M}_1$ message into two independent sub-message sets $\mathcal{M}_{11}$ and $\mathcal{M}_{12}$ so that $R_1 = R_{11} + R_{12}$.

*Codebook generation:* First randomly and independently generate $2^{n(R_0'+R_0)}$ sequences $u(m_0', m_0), m_0' \in \{1,2,...,2^{nR_0'}\}$, $m_0 \in \{1,2,...,2^{nR_0}\}$, each one i.i.d according to $\prod_{i=1}^{n} p(u_i)$ and then randomly throw them into $2^{nR_0}$ bins. It is clear that we have $2^{nR_0'}$ sequences in each bin.

Now for each $u(m_0', m_0)$, randomly and independently generate $2^{n(R_{11}'+R_{11})}$ sequences $v(m_0', m_0, m_{11}', m_{11}), m_{11}' \in \{1,...,2^{nR_{11}'}\}$, $m_{11} \in \{1,...,2^{nR_{11}}\}$ each one i.i.d according to $\prod_{i=1}^{n} p_{V|U}(v_i|u_i(m_0', m_0))$, and randomly throw them into $2^{nR_{11}}$ bins.

Now for each sequence $v(m_0', m_0, m_{11}', m_{11})$, randomly and independently generate $2^{n(R_{12}'+R_{12})}$ sequences $x(m_0', m_0, m_{11}', m_{11}, m_{12}', m_{12})$ each one i.i.d according to $\prod_{i=1}^{n} p_{X|U,V}(x_i|v_i, u_i) = \prod_{i=1}^{n} p_{X|V}(x_i|v_i)$. Then randomly throw them into $2^{nR_{12}}$ bins. Then provide the transmitter and all the receivers with bins and their codewords.

***Encoding:*** We are given the side information $s$ and the message pair $(m_0, m_1)$. Indeed, our messages are bin indices. We find $m_{11}$, and $m_{12}$. Now in the bin $m_0$ of $u$ sequences look for a $m_0'$ such that $(u(m_0', m_0), s) \in A_\epsilon^{(n)}$, i.e. the sequence $u$ that is jointly typical with the $s$ given where definitions of typical sequences are given in section II-B. Then in the bin $m_{11}$ of $v$ sequences look for some $m_{11}'$ such that

$$(u(m_0', m_0), v(m_0', m_0, m_{11}', m_{11}), s) \in A_\epsilon^{(n)}$$

Now in the bin $m_{12}$ of $x$ sequences look for some $m_{12}'$ such that

$$(u(m_0', m_0), v(m_0', m_0, m_{11}', m_{11}),$$
$$x(m_0', m_0, m_{11}', m_{11}, m_{12}', m_{12}), s) \in A_\epsilon^{(n)}$$

We send the found $x$ sequence. Before bumping into decoding, assume that the correct indices are found through the encoding procedure, i.e. $m_0' = M_0', m_{11}' = M_{11}'$, and $m_{12}' = M_{12}'$.

***Decoding:*** Since the messages are uniformly distributed over their respective ranges, we can assume, without loss of generality, that the tuple $(m_0, m_{11}, m_{12}) = (1,1,1)$ is sent.

The second receiver $Y_2$ receives $y_2$ thus having the following error events

$$E_{21} = \left\{ (u(M_0', 1), y_2) \notin A_\epsilon^{(n)} \right\}$$

$$E_{22} = \{ (u(m_0', m_0), y_2) \in A_\epsilon^{(n)} \text{ for some } m_0 \neq 1$$
$$\text{and } m_0' \neq M_0' \}$$

***Remark 2.1:*** the following error event

$$E_{23} = \left\{ (u(M_0', m_0), y_1) \in A_\epsilon^{(n)} \text{ for some } m_0 \neq 1 \right\}$$

leads us to a redundant inequality.

Now by the weak law of large numbers (WLLN) [25], $p(E_{21}) \leq \epsilon, \forall \epsilon > 0$ as $n \to \infty$. For the second error event we have

$$p(E_{22}) = \sum_{m_0, m_0'} \sum_{A_\epsilon^{(n)}} p(u) p(y_2) \leq 2^{n(R_0 + R_0')} 2^{n(H(U, Y_2) + \epsilon)}$$
$$2^{-n(H(U) - \epsilon)} 2^{-n(H(Y_2) - \epsilon)} = 2^{-n(I(U;Y_2) - 3\epsilon - R_0 - R_0')}$$

We see that $\forall \epsilon > 0$, $p(E_{22}) \leq \epsilon$ as $n \to \infty$ provided that

$$R_0 + R_0' \leq I(U; Y_2) - 3\epsilon \quad (6)$$

The first receiver $Y_1$ receives $y_1$ and needs to decode both $m_0$ and $m_1$. Therefore, the error events are

$$E_{11} = \{ (u(M_0', 1), v(M_0', 1, M_{11}', 1),$$
$$x(M_0', 1, M_{11}', 1, M_{12}', 1), y_1) \notin A_\epsilon^{(n)} \}$$

$$E_{12} = \{ (u(M_0', 1), v(M_0', 1, M_{11}', 1),$$
$$x(M_0', 1, M_{11}', 1, m_{12}', m_{12}), y_1) \in A_\epsilon^{(n)}$$
$$\text{for some } m_{12} \neq 1 \text{ and } m_{12}' \neq M_{12}' \}$$

$$E_{13} = \{ (u(M_0', 1), v(M_0', 1, m_{11}', m_{11}),$$
$$x(M_0', 1, m_{11}', m_{11}, m_{12}', m_{12}), y_1) \in A_\epsilon^{(n)}$$
$$\text{for some } m_{1i} \neq 1 \text{ and } m_{1i}' \neq M_{1i}', i = 1,2 \}$$

$$E_{14} = \{ (u(m_0', m_0), v(m_0', m_0, m_{11}', m_{11}),$$
$$x(m_0', m_0, m_{11}', m_{11}, m_{12}', m_{12}), y_1) \in A_\epsilon^{(n)}$$
$$\text{for some } m_0 \neq 1 \text{ and } m_0' \neq M_0' \text{ and some}$$
$$m_{1i} \neq 1 \text{ and } m_{1i}' \neq M_{1i}', i = 1,2 \}$$

The first receiver's probability of error can be arbitrarily made small provided that

$$R_{12} + R_{12}' \leq I(X; Y_1|V) - 6\epsilon \quad (7)$$
$$R_{11} + R_{11}' + R_{12} + R_{12}' \leq I(X; Y_1|U) - 6\epsilon \quad (8)$$
$$R_0 + R_0' + R_{11} + R_{11}' + R_{12} + R_{12}' \leq I(X; Y_1) - 5\epsilon \quad (9)$$

The third receiver $Y_3$ receives $y_3$ and needs to decode only the common message indirectly by decoding the message $m_{11}$. The error events are

$$E_{31} = \left\{ (u(M_0', 1), v(M_0', 1, M_{11}', 1), y_3) \notin A_\epsilon^{(n)} \right\}$$

$$E_{32} = \{ (u(M_0', 1), v(M_0', 1, m_{11}', m_{11}), y_3) \in A_\epsilon^{(n)} \text{ for}$$
$$\text{some } m_{11} \neq 1 \text{ and } m_{11}' \neq M_{11}' \}$$

$$E_{33} = \{ (u(m_0', m_0), v(m_0', m_0, m_{11}', m_{11}), y_3) \in A_\epsilon^{(n)} \text{ for}$$
$$\text{some } m_0 \neq 1, m_{11} \neq 1, m_0' \neq M_0', \text{and } m_{11}' \neq M_{11}' \}$$

Again by using WLLN and AEP, we see that the third receiver's error probabilities can be arbitrarily made small as $n \to \infty$ provided that

$$R_0 + R_0' + R_{11} + R_{11}' \leq I(V; Y_3) - 3\epsilon \quad (10)$$

Using Gel'fand-Pinsker coding we see that the encoders can choose the proper $m_0', m_{11}'$, and $m_{12}'$ indices with vanishing probability of error provided that for every $\epsilon > 0$ and sufficiently large n

$$R_0' \geq I(U; S) + 2\epsilon \quad (11)$$
$$R_{11}' \geq I(V; S|U) + 2\epsilon \quad (12)$$
$$R_{12}' \geq I(X; S|V) + 2\epsilon \quad (13)$$

Now combining (6) – (10) and (11) – (13) and noting that

$$I(V; S|U) + I(U; S) = I(VU; S) \quad (14)$$

and using Fourier-Motzkin procedure afterwards to eliminate $R_{11}$ and $R_{12}$, we obtain (5) as an achievable rate region for MBC with SI. ∎

***Remark 2.2:*** After Fourier-Motzkin elimination, we get the following inequality that seems to have to be added to (5)

$$R_0 + R_1 \leq I(X; Y_1) - I(UV; S) - I(X; S|V).$$

But this is actually a redundant inequality since

1. If $I(U; Y_2) - I(U; S) \leq I(V; Y_3) - I(UV; S)$,

then we have

$$R_0 + R_1 \leq I(U;Y_2) - I(U;S) + I(X;Y_1|U)$$
$$-I(V;S|U) - I(X;S|V) \leq I(U;Y_1) + I(X;Y_1|U)$$
$$-I(UV;S) - I(X;S|V).$$

2. If $I(V;Y_3) - I(UV;S) \leq I(U;Y_2) - I(U;S)$, then we have

$$R_0 + R_1 \leq I(V;Y_3) - I(UV;S) + I(X;Y_1|U)$$
$$-I(V;S|U) - I(X;S|V) \leq I(U;Y_2) - I(U;S)$$
$$+I(X;Y_1|U) - I(V;S|U) - I(X;S|V) \leq I(X;Y_1)$$
$$-I(UV;S) - I(X;S|V).$$

***Theorem 2:*** The capacity region of MBC with SI non-causally available at the transmitter and at all the receivers with one deterministic component, i.e. when there exists a deterministic function $f$ such that $Y_3 = f(X,S)$, is the set of all rate pairs $(R_0, R_1)$ such that

$$R_0 \leq \min\{I(U;Y_2|S), H(Y_3|S)\}$$
$$R_1 \leq I(X;Y_1|U,S) \quad (15)$$
$$R_0 + R_1 \leq H(Y_3|S) + I(X;Y_1|Y_3,S)$$

for some

$$p(u,v,s,x,y_1,y_2,y_3) = p(s)p(u|s)p(v|u,s)$$
$$p(x|v,s)p(y_1|x,s)I(y_3 = f(x,s))p(y_2|y_1)$$

where I(.) is the identity function.

***Proof:***

*Achievability:* Setting $Y_k = (Y_k, S), k = 1,2,3$, in (5) we see that

$$R_0 \leq I(U;Y_2,S) - I(U;S) = I(U;Y_2|S)$$

$$R_0 \leq I(V;Y_3,S) - I(UV;S) = I(UV;Y_3,S) - I(UV;S)$$
$$= I(UV;Y_3|S) = I(V;Y_3|S)$$

$$R_1 \leq I(X;Y_1,S|U) - I(V;S|U) - I(X;S|V)$$
$$= I(V,X;Y_1,S|U) - I(V;S|U) - I(X;S|V)$$
$$= I(V,X;S|U) + I(V,X;Y_1|U,S) - I(V;S|U) - I(X;S|V)$$
$$= I(X;Y_1|U,S) + I(X;S|U,V) - I(X;S|V) = I(X;Y_1|U,S)$$

Notice that the last equality follows from the fact that

$$I(X;S|U,V) = H(X|U,V) - H(X|U,V,S)$$
$$= H(X|V) - H(X|V,S) = I(X;S|V)$$

$$R_0 + R_1 \leq I(U,V;Y_3,S) + I(X;Y_1,S|V) - I(X;S|V)$$
$$-I(U,V;S) = I(U,V;Y_3|S) + I(X;Y_1|V,S)$$
$$= I(V;Y_3|S) + I(X;Y_1|V,S)$$

Therefore (5) is reduced to

$$R_0 \leq \min\{I(U;Y_2|S), I(V;Y_3|S)\}$$
$$R_1 \leq I(X;Y_1|U,S) \quad (16)$$
$$R_0 + R_1 \leq I(V;Y_3|S) + I(X;Y_1|V,S)$$

Notice that we always have the following Markov chain from (2)

$$V \to (X,S) \to Y_3$$

and since $Y_3 = f(X,S)$, we can set $V = Y_3$ in (16) to obtain (15).

*Converse:* By the memorylessness of the channel we have

$$(M_0, M_1, Y_1^{i-1}, Y_2^{i-1}, Y_3^{i-1}, S_{i+1}^{(n)}, S^{i-1}) \to (X_i, S_i) \to$$
$$(Y_{1i}, Y_{2i}, Y_{3i}). \quad (17)$$

From (17) we obtain

$$(M_0, M_1, Y_1^{i-1}, Y_3^{i-1}, S_{i+1}^{(n)}, S^{i-1}) \to (X_i, S_i, Y_2^{i-1}) \to$$
$$(Y_{1i}, Y_{2i}, Y_{3i}) \quad (18)$$

due to the centripetal property of first order Markov chains which is itself established by the non-negativity of mutual information.

Now assume that the code $(n, 2^{nR_0}, 2^{nR_1}, \epsilon)$ is $\epsilon$-achievable for the MBC with SI with one deterministic component, therefore using Fano's inequality we have

$$nR_0 = H(M_0) = H(M_0|S^n) = H(M_0|S^n, Y_2^n) + I(M_0; Y_2^n|S^n)$$
$$\leq H(M_0|Y_2^n) + \sum_{i=1}^{n} I(M_0; Y_{2i}|Y_2^{i-1}, S^{i-1}, S_{i+1}^{(n)}, S_i)$$
$$\leq n\epsilon_{02n} + \sum_{i=1}^{n} I(M_0, Y_1^{i-1}, Y_2^{i-1}, S^{i-1}, S_{i+1}^{(n)}; Y_{2i}|S_i)$$
$$= n\epsilon_{02n} + \sum_{i=1}^{n} I(U_i; Y_{2i}|S_i)$$

where $U_i \triangleq (M_0, Y_1^{i-1}, Y_2^{i-1}, S^{i-1}, S_{i+1}^{(n)})$. We also have

$$nR_0 = H(M_0) = H(M_0|M_1, S^n)$$
$$= H(M_0|M_1, S^n, Y_3^n) + I(M_0; Y_3^n|M_1, S^n)$$
$$\leq H(M_0|Y_3^n) + \sum_{i=1}^{n} I(M_0; Y_{3i}|M_1, Y_3^{i-1}, S^{i-1}, S_{i+1}^{(n)}, S_i)$$
$$\stackrel{(a)}{=} n\epsilon_{03n} + \sum_{i=1}^{n} H(Y_{3i}|M_1, Y_3^{i-1}, S^{i-1}, S_{i+1}^{(n)}, S_i)$$
$$\leq n\epsilon_{03n} + \sum_{i=1}^{n} H(Y_{3i}|S_i)$$

where (a) follows from the fact that $Y_3$ is a function of $(M_0, M_1, S^n)$. For the bound on $R_1$ we have

$$nR_1 = H(M_1) = H(M_1|M_0, S^n)$$
$$= H(M_1|M_0, S^n, Y_1^n) + I(M_1; Y_1^n|M_0, S^n)$$
$$\leq \sum_{i=1}^{n} I(M_1; Y_{1i}|M_0, Y_1^{i-1}, S^{i-1}, S_{i+1}^{(n)}, S_i) + H(M_1|Y_1^n)$$
$$\stackrel{(a)}{\leq} \sum_{i=1}^{n} I(X_i, S_i, Y_2^{i-1}; Y_{1i}|M_0, Y_1^{i-1}, S^{i-1}, S_{i+1}^{(n)}, S_i) + n\epsilon_{11n}$$
$$= \sum_{i=1}^{n} I(Y_2^{i-1}; Y_{1i}|M_0, Y_1^{i-1}, S^{i-1}, S_{i+1}^{(n)}, S_i) + n\epsilon_{11n}$$

$$+ \sum_{i=1}^{n} I(X_i; Y_{1i} | M_0, Y_1^{i-1}, Y_2^{i-1}, S^{i-1}, S_{i+1}^{(n)}, S_i)$$

$$\stackrel{(b)}{=} \sum_{i=1}^{n} I(X_i; Y_{1i} | M_0, Y_1^{i-1}, Y_2^{i-1}, S^{i-1}, S_{i+1}^{(n)}, S_i) + n\epsilon_{11n}$$

$$= \sum_{i=1}^{n} I(X_i; Y_{1i} | U_i, S_i) + n\epsilon_{11n}.$$

where (a) follows from (18) and (b) follows from the Markov chain $Y_2^{i-1} \to (M_0, Y_1^{i-1}, S^{i-1}, S_{i+1}^{(n)}, S_i) \to Y_{1i}$ due to the degradedness of $Y_2$ than to $Y_1$.

The last bound on $R_0 + R_1$ is proved as follows

$$n(R_0 + R_1) = H(M_0, M_1 | S^n)$$
$$= H(M_0, M_1 | S^n, Y_1^n, Y_3^n) + I(M_0, M_1; Y_1^n, Y_3^n | S^n)$$
$$\leq n\epsilon_{2n} + \sum_{i=1}^{n} I(M_0, M_1; Y_{1i}, Y_{3i} | S^{i-1}, S_{i+1}^{(n)}, S_i, Y_1^{i-1}, Y_3^{i-1})$$
$$= n\epsilon_{2n} + \sum_{i=1}^{n} H(Y_{1i}, Y_{3i} | S^{i-1}, S_{i+1}^{(n)}, S_i, Y_1^{i-1}, Y_3^{i-1})$$
$$- \sum_{i=1}^{n} H(Y_{1i}, Y_{3i} | M_0, M_1, S^{i-1}, S_{i+1}^{(n)}, S_i, Y_1^{i-1}, Y_3^{i-1})$$
$$\leq n\epsilon_{2n} + \sum_{i=1}^{n} H(Y_{1i}, Y_{3i} | S_i)$$
$$- \sum_{i=1}^{n} H(Y_{1i} | M_0, M_1, S^{i-1}, S_{i+1}^{(n)}, S_i, Y_1^{i-1}, Y_3^{i-1})$$
$$\stackrel{(a)}{=} n\epsilon_{2n} + \sum_{i=1}^{n} H(Y_{3i} | S_i) + H(Y_{1i} | Y_{3i}, S_i)$$
$$- \sum_{i=1}^{n} H(Y_{1i} | M_0, M_1, S^{i-1}, S_{i+1}^{(n)}, S_i, Y_1^{i-1}, Y_3^{i-1}, Y_{3i})$$
$$= n\epsilon_{2n} + \sum_{i=1}^{n} H(Y_{3i} | S_i)$$
$$+ \sum_{i=1}^{n} I(M_0, M_1, S^{i-1}, S_{i+1}^{(n)}, Y_1^{i-1}, Y_3^{i-1}; Y_{1i} | Y_{3i}, S_i)$$
$$\stackrel{(b)}{\leq} n\epsilon_{2n} + \sum_{i=1}^{n} H(Y_{3i} | S_i) + I(X_i, S_i; Y_{1i} | Y_{3i}, S_i)$$
$$= n\epsilon_{2n} + \sum_{i=1}^{n} H(Y_{3i} | S_i) + I(X_i; Y_{1i} | Y_{3i}, S_i).$$

where (a) follows from the chain rule for entropy and the fact that $Y_{3i} = f(M_0, M_1, S^n)$ and (b) follows from (17).

Therefore the proof of the converse part is immediate using the standard time-sharing schemes and the Theorem's proof is complete. ∎

***Proposition 2.1:*** The capacity region of MBC with SI non-causally available at the transmitter and at all the receivers with two deterministic components, i.e. when there exist two deterministic functions $f_1$, and $f_3$ such that $Y_1 = f_1(X, S)$, and $Y_3 = f_3(X, S)$, is the set of all rate pairs $(R_0, R_1)$ such that

$$\begin{aligned} R_0 &\leq \min\{I(U; Y_2|S), H(Y_3|S)\} \\ R_1 &\leq H(Y_1|U, S) \\ R_0 + R_1 &\leq H(Y_1, Y_3|S) \end{aligned} \tag{19}$$

for some

$$p(u, v, s, x, y_1, y_2, y_3) = p(s)p(u|s)p(v|u,s)$$
$$p(x|v,s)I(y_1 = f_1(x,s))I(y_3 = f_3(x,s))p(y_2|y_1)$$

*Proof:*

*Achievability:* We can obtain (19) by setting $X = Y_1$ and $V = Y_3$ in (16).

*Converse:* Bound on $R_0$ is the same as Theorem 2. Now using Fano's inequality we have

$$nR_1 \leq n\epsilon_{11n} + \sum_{i=1}^{n} I(M_1; Y_{1i} | M_0, Y_1^{i-1}, S^{i-1}, S_{i+1}^{(n)}, S_i)$$
$$\leq n\epsilon_{11n} + \sum_{i=1}^{n} I(M_1, X_i, Y_2^{i-1}; Y_{1i} | M_0, Y_1^{i-1}, S^{i-1}, S_{i+1}^{(n)}, S_i)$$
$$\stackrel{(a)}{=} n\epsilon_{11n} + \sum_{i=1}^{n} I(M_1, X_i; Y_{1i} | M_0, Y_1^{i-1}, Y_2^{i-1}, S^{i-1}, S_{i+1}^{(n)}, S_i)$$
$$\stackrel{(b)}{=} n\epsilon_{11n} + \sum_{i=1}^{n} H(Y_{1i} | M_0, Y_1^{i-1}, Y_2^{i-1}, S^{i-1}, S_{i+1}^{(n)}, S_i)$$
$$= n\epsilon_{11n} + \sum_{i=1}^{n} H(Y_{1i} | U_i, S_i),$$

where (a) follows from $Y_2^{i-1} \to (M_0, Y_1^{i-1}, S^n) \to Y_{1i}$, (b) follows from the fact that $Y_{1i}$ is a function of $(M_0, M_1, S^n)$.

For sum of the rates we have

$$n(R_0 + R_1) = H(M_0, M_1 | S^n)$$
$$= H(M_0, M_1 | S^n, Y_1^n, Y_3^n) + I(M_0, M_1; Y_1^n, Y_3^n | S^n)$$
$$\leq n\epsilon_{0n} + \sum_{i=1}^{n} I(M_0, M_1; Y_{1i}, Y_{3i} | Y_1^{i-1}, Y_3^{i-1}, S^n)$$
$$= n\epsilon_{0n} + \sum_{i=1}^{n} H(Y_{1i}, Y_{3i} | Y_1^{i-1}, Y_3^{i-1}, S^n)$$
$$\leq n\epsilon_{0n} + \sum_{i=1}^{n} H(Y_{1i}, Y_{3i} | S_i). \quad \blacksquare$$

***Proposition 2.2:*** The capacity region of deterministic MBC with SI non-causally available at the transmitter and at all the receivers, i.e. when there exist three deterministic functions $f_1, f_2$, and $f_3$ such that $Y_1 = f_1(X, S), Y_2 = f_2(X, S)$, and $Y_3 = f_3(X, S)$, is the set of all rate pairs $(R_0, R_1)$ such that

$$\begin{aligned} R_0 &\leq \min\{H(Y_2|S), H(Y_3|S)\} \\ R_1 &\leq H(Y_1|Y_2, S) \\ R_0 + R_1 &\leq H(Y_1, Y_3|S) \end{aligned} \tag{20}$$

for some

$$p(u,v,s,x,y_1,y_2,y_3) = p(s)p(u|s)p(v|u,s)$$
$$p(x|v,s)I(y_1 = f_1(x,s))I(y_2 = f_2(x,s))I(y_3 = f_3(x,s))$$

*Proof:*

*Achievability:* The achievability part is immediate if we set $U = Y_2$ in (19).

*Converse:*

$$nR_0 = H(M_0|S^n, Y_2^n) + I(M_0; Y_2^n|S^n)$$
$$\leq n\epsilon_{02n} + \sum_{i=1}^{n} I(M_0; Y_{2i}|Y_2^{i-1}, S^n)$$
$$= n\epsilon_{02n} + \sum_{i=1}^{n} H(Y_{2i}|Y_2^{i-1}, S^n) \leq n\epsilon_{02n} + \sum_{i=1}^{n} H(Y_{2i}|S_i).$$

$$nR_1 = H(M_1|M_0, S^n) \leq n\epsilon_{11n} + \sum_{i=1}^{n} I(M_1; Y_{1i}|M_0, Y_1^{i-1}, S^n)$$
$$= n\epsilon_{11n} + \sum_{i=1}^{n} H(Y_{1i}|M_0, Y_1^{i-1}, S^n)$$
$$\stackrel{(a)}{=} n\epsilon_{11n} + \sum_{i=1}^{n} H(Y_{1i}|M_0, Y_1^{i-1}, S^n, Y_{2i})$$
$$\leq n\epsilon_{11n} + \sum_{i=1}^{n} H(Y_{1i}|Y_{2i}, S_i).$$

where (a) follows from the fact that $Y_{2i}$ is function of $(M_0, S^n)$. ∎

## IV. THREE-RECEIVER LESS NOISY BROADCAST CHANNEL WITH SIDE INFORMATION

Define $\mathcal{P}^*$ as the collection of all random variables $(U, V, S, X, Y_1, Y_2, Y_3)$ with finite alphabets such that

$$p(u,v,s,x,y_1,y_2,y_3) =$$
$$p(s)p(u|s)p(v|u,s)p(x|v,s)p(y_1,y_2,y_3|x,s) \quad (21)$$

*Theorem 3:* A rate triple $(R_1, R_2, R_3)$ is achievable for 3-receiver less noisy broadcast channel with SI non-causally available at the transmitter provided that

$$R_1 \leq I(X; Y_1|V) - I(X; S|V)$$
$$R_2 \leq I(V; Y_2|U) - I(V; S|U) \quad (22)$$
$$R_3 \leq I(U; Y_3) - I(U; S)$$

for some joint distribution on $\mathcal{P}^*$.

*Corollary 3.1:* By setting $S \equiv \emptyset$ in the above rate region, it reduces to the capacity region of 3-receiver less noisy BC given in [13].

Proof: The proof uses Cover's superposition [28] and Gel'fand-Pinsker random binning coding [15] procedures along with Nair's indirect decoding and is similar to the proof of Theorem 1 and thus only an outline is provided.

Fix n and a distribution on $\mathcal{P}^*$.

Again note that side information is distributed i.i.d according to

$$p(s) = \prod_{i=1}^{n} p(s_i)$$

Randomly and independently generate $2^{n(R_3'+R_3)}$ sequences $\boldsymbol{u}(m_3', m_3)$, each distributed i.i.d according to $\prod_{i=1}^{n} p(u_i)$ and randomly throw them into $2^{nR_3}$ bins.

For each $\boldsymbol{u}(m_3', m_3)$, randomly and independently generate $2^{n(R_2'+R_2)}$ sequences $\boldsymbol{v}(m_3', m_3, m_2', m_2)$, each distributed i.i.d according to $\prod_{i=1}^{n} p_{V|U}(v_i|u_i)$ and randomly throw them into $2^{nR_2}$ bins.

Now for each generated $\boldsymbol{v}(m_3', m_3, m_2', m_2)$, randomly and independently generate $2^{n(R_1'+R_1)}$ sequences $\boldsymbol{x}(m_3', m_3, m_2', m_2, m_1', m_1)$ each one distributed i.i.d according to $\prod_{i=1}^{n} p_{X|V}(x_i|v_i)$ and randomly throw them into $2^{nR_1}$ bins.

Encoding is succeeded with small probability of error provided that

$$R_3' \geq I(U; S) \quad (23)$$
$$R_2' \geq I(V; S|U) \quad (24)$$
$$R_1' \geq I(X; S|V) \quad (25)$$

and decoding is succeeded if

$$R_3 + R_3' \leq I(U; Y_3) \quad (26)$$
$$R_2 + R_2' \leq I(V; Y_2|U) \quad (27)$$
$$R_1 + R_1' \leq I(X; Y_1|V) \quad (28)$$

Now combining (23), (24), and (25) with (26), (27) and (28) gives us (22). ∎

*Theorem 4:* The capacity region of the three-receiver less noisy BC with SI non-causally available at the transmitter and at the receivers is the set of all rate triples $(R_1, R_2, R_3)$ such that

$$\begin{aligned} R_1 &\leq I(X; Y_1|VS) \\ R_2 &\leq I(V; Y_2|US) \\ R_3 &\leq I(U; Y_3|S) \end{aligned} \quad (29)$$

*Proof:*

*Achievability:* The direct part of the proof is achieved if one sets $Y_k = (Y_k, S), k = 1,2,3$, in (22).

*Converse:* The converse part uses an extension of lemma 1 in [13].

*Lemma 1:* [13] Let the channel $X \to Y$ be less noisy than the channel $X \to Z$. Consider $(M, S^n)$ to be any random vector such that

$$(M, S^n) \to X^n \to (Y^n, Z^n)$$

forms a Markov chain. Then

1. $I(Y^{i-1}; Z_i | M, S^n) \geq I(Z^{i-1}; Z_i | M, S^n)$
2. $I(Y^{i-1}; Y_i | M, S^n) \geq I(Z^{i-1}; Y_i | M, S^n)$

*Proof:* First of all note that since the channel is memoryless we have

$$(M_1, M_2, M_3, Y_1^{i-1}, Y_2^{i-1}, Y_3^{i-1}, S^{i-1}, S_{i+1}^n) \to (X_i, S_i) \to (Y_{1i}, Y_{2i}, Y_{3i})$$

Just like [13], for any $1 \leq r \leq i-1$

$$I(Z^{r-1}, Y_r^{i-1}; Y_i | M, S^n)$$
$$= I(Z^{r-1}, Y_{r+1}^{i-1}; Y_i | M, S^n)$$
$$+ I(Y_r; Y_i | M, S^n, Z^{r-1}, Y_{r+1}^{i-1})$$
$$\geq I(Z^{r-1}, Y_{r+1}^{i-1}; Y_i | M, S^n)$$
$$+ I(Z_r; Y_i | M, S^n, Z^{r-1}, Y_{r+1}^{i-1})$$
$$= I(Z^r, Y_{r+1}^{i-1}; Y_i | M, S^n)$$

where the inequality follows from the memorylessness of the channel and the fact that $Y$ is less noisy than $Z$, i.e.

$$I(Y_r; Y_i | M, S^n, Z^{r-1}, Y_{r+1}^{i-1}) \geq I(Z_r; Y_i | M, S^n, Z^{r-1}, Y_{r+1}^{i-1}).$$

Proof of the second part follows the same lines as the first part with negligible variations. ∎

Now we stick to the proof of the converse

$$nR_3 = H(M_3 | S^n) = H(M_3 | S^n, Y_3^n) + I(M_3; Y_3^n | S^n)$$
$$\leq H(M_3 | Y_3^n) + I(M_3; Y_3^n | S^n)$$
$$\leq n\epsilon_{3n} + \sum_{i=1}^n I(M_3; Y_{3i} | S^n, Y_3^{i-1})$$
$$\leq n\epsilon_{3n} + \sum_{i=1}^n I(M_3; Y_{3i} | S^{i-1}, S_i, S_{i+1}^n, Y_3^{i-1})$$
$$\leq n\epsilon_{3n} + \sum_{i=1}^n I(M_3, S^{i-1}, S_{i+1}^n, Y_3^{i-1}; Y_{3i} | S_i)$$
$$\leq n\epsilon_{3n} + \sum_{i=1}^n I(M_3, S^{i-1}, S_{i+1}^n, Y_2^{i-1}; Y_{3i} | S_i)$$
$$= n\epsilon_{3n} + \sum_{i=1}^n I(U_i; Y_{3i} | S_i).$$

where $U_i \triangleq (M_3, S^{i-1}, S_{i+1}^n, Y_2^{i-1})$ and the last inequality follows from lemma 1.

$$nR_2 = H(M_2 | M_3, S^n) = H(M_2 | M_3, S^n, Y_2^n)$$
$$+ I(M_2; Y_2^n | M_3, S^n) \leq H(M_2 | Y_2^n) + I(M_2; Y_2^n | M_3, S^n)$$
$$\leq n\epsilon_{2n} + \sum_{i=1}^n I(M_2; Y_{2i} | M_3, S^{i-1}, S_i, S_{i+1}^n, Y_2^{i-1}) = n\epsilon_{2n}$$
$$+ \sum_{i=1}^n I(M_2, M_3, S^{i-1}, S_{i+1}^n, Y_2^{i-1}; Y_{2i} | M_3, S^{i-1}, S_{i+1}^n, Y_2^{i-1}, S_i)$$
$$= n\epsilon_{2n} + \sum_{i=1}^n I(V_i; Y_{2i} | U_i, S_i),$$

where $V_i \triangleq (M_2, M_3, S^{i-1}, S_{i+1}^n, Y_2^{i-1})$. It is clear that for the given choice of $U_i$ and $V_i$, we have the Markov chain (2) satisfied for the channel is assumed to be memoryless.

$$nR_1 = H(M_1 | S^n, M_2, M_3) = H(M_1 | M_2, M_3, S^n, Y_1^n)$$
$$+ I(M_1; Y_1^n | M_2, M_3, S^n)$$
$$\leq H(M_1 | Y_1^n) + I(M_1; Y_1^n | M_2, M_3, S^n)$$
$$\leq n\epsilon_{1n} + \sum_{i=1}^n I(M_1; Y_{1i} | M_2, M_3, S^{i-1}, S_i, S_{i+1}^n, Y_1^{i-1})$$
$$\stackrel{(a)}{\leq} n\epsilon_{1n} + \sum_{i=1}^n I(X_i; Y_{1i} | M_2, M_3, S^{i-1}, S_i, S_{i+1}^n, Y_1^{i-1})$$
$$= n\epsilon_{1n} + \sum_{i=1}^n I(X_i; Y_{1i} | M_2, M_3, S^{i-1}, S_i, S_{i+1}^n)$$
$$- \sum_{i=1}^n I(Y_1^{i-1}; Y_{1i} | M_2, M_3, S^{i-1}, S_i, S_{i+1}^n)$$
$$\stackrel{(b)}{\leq} n\epsilon_{1n} + \sum_{i=1}^n I(X_i; Y_{1i} | M_2, M_3, S^{i-1}, S_i, S_{i+1}^n)$$
$$- \sum_{i=1}^n I(Y_2^{i-1}; Y_{1i} | M_2, M_3, S^{i-1}, S_i, S_{i+1}^n)$$
$$= n\epsilon_{1n} + \sum_{i=1}^n I(X_i; Y_{1i} | M_2, M_3, S^{i-1}, S_{i+1}^n, Y_2^{i-1}, S_i)$$
$$= n\epsilon_{1n} + \sum_{i=1}^n I(X_i; Y_{1i} | U_i, S_i),$$

where (a) follows from the memorylessness of the channel and (b) follows from lemma 1.

Now using the standard time-sharing scheme, we can easily conclude that any achievable rate triple for the three-receiver less noisy BC with SI non-causally available at the transmitter and at the receivers, must satisfy (29) and the proof is complete. ∎

*Proposition 3.1:* The capacity region of the three-receiver less noisy BC with SI non-causally available at the transmitter and at all the receivers with one deterministic component, i.e. when there exists a function $f_1$ such that $Y_1 = f_1(X, S)$, is the set of all rate triples $(R_1, R_2, R_3)$ such that

$$\begin{aligned} R_1 &\leq H(Y_1 | V, S) \\ R_2 &\leq I(V; Y_2 | U, S) \\ R_3 &\leq I(U; Y_3 | S) \end{aligned} \quad (30)$$

*Achievability:* By setting $X = Y_1$ in (29), (30) is obtained.

*Converse:* Bounds on $R_2$ and $R_3$ and the choice of auxiliary random variables $U_i$ and $V_i$ are the same as that of Theorem 4. Bound on $R_1$, though, is:

$$nR_1 = H(M_1 | M_2, M_3, S^n, Y_1^n) + I(M_1; Y_1^n | M_2, M_3, S^n)$$
$$= n\epsilon_{1n} + \sum_{i=1}^n I(M_1; Y_{1i} | M_2, M_3, S^n, Y_1^{i-1})$$

$$\leq n\epsilon_{1n} + \sum_{i=1}^{n} I(X_i; Y_{1i}|M_2, M_3, S^n, Y_1^{i-1})$$

$$= n\epsilon_{1n} + \sum_{i=1}^{n} I(X_i; Y_{1i}|M_2, M_3, S^n) - I(Y_1^{i-1}; Y_{1i}|M_2, M_3, S^n)$$

$$\leq n\epsilon_{1n} + \sum_{i=1}^{n} I(X_i; Y_{1i}|M_2, M_3, S^n) - I(Y_2^{i-1}; Y_{1i}|M_2, M_3, S^n)$$

$$= n\epsilon_{1n} + \sum_{i=1}^{n} I(X_i; Y_{1i}|M_2, M_3, S^n, Y_2^{i-1})$$

$$= n\epsilon_{1n} + \sum_{i=1}^{n} H(Y_{1i}|M_2, M_3, S^{i-1}, S_{i+1}^n, Y_2^{i-1}, S_i)$$

$$= n\epsilon_{1n} + \sum_{i=1}^{n} H(Y_{1i}|V_i, S_i)$$

where again $V_i \triangleq (M_2, M_3, S^{i-1}, S_{i+1}^n, Y_2^{i-1})$. ∎

*Proposition 3.2:* The capacity region of the three-receiver less noisy BC with SI non-causally available at the transmitter and at all the receivers with two deterministic components, i.e. when there exist two deterministic functions $f_1$, and $f_2$ such that $Y_1 = f_1(X, S)$ and $Y_2 = f_2(X, S)$, is the set of all rate triples $(R_1, R_2, R_3)$ such that

$$\begin{aligned} R_1 &\leq H(Y_1|Y_2 S) \\ R_2 &\leq H(Y_2|U, S) \\ R_3 &\leq I(U; Y_3|S) \end{aligned} \quad (31)$$

*Achievability:* The direct part is proved by setting $X = Y_1$, and $V = Y_2$ in (29).

*Converse:* The bound on $R_3$ is the same as in Theorem 4. For other bounds using Fano's inequality we have

$$nR_2 \leq n\epsilon_{2n} + \sum_{i=1}^{n} I(M_2; Y_{2i}|M_3, S^{i-1}, S_i, S_{i+1}^n, Y_2^{i-1})$$

$$\stackrel{(a)}{=} n\epsilon_{2n} + \sum_{i=1}^{n} H(Y_{2i}|M_3, S^{i-1}, S_{i+1}^n, Y_2^{i-1}, S_i)$$

$$= n\epsilon_{2n} + \sum_{i=1}^{n} H(Y_{2i}|U_i, S_i),$$

$$nR_1 \leq n\epsilon_{1n} + \sum_{i=1}^{n} I(M_1; Y_{1i}|M_2, M_3, S^n, Y_1^{i-1})$$

$$= n\epsilon_{1n} + \sum_{i=1}^{n} H(Y_{1i}|M_2, M_3, S^n, Y_1^{i-1})$$

$$\stackrel{(b)}{=} n\epsilon_{1n} + \sum_{i=1}^{n} H(Y_{1i}|M_2, M_3, S^n, Y_1^{i-1}, Y_{2i})$$

$$\leq n\epsilon_{1n} + \sum_{i=1}^{n} H(Y_{1i}|Y_{2i}, S_i).$$

where (a) and (b) both follows from the fact that $Y_{2i}$ is a function of $(M_2, S^n)$. ∎

*Proposition 3.3:* The capacity region of deterministic three-receiver less noisy BC with SI non-causally available at the transmitter and at all the receivers is the set of all rate triples $(R_1, R_2, R_3)$ such that

$$\begin{aligned} R_1 &\leq H(Y_1|Y_2, S) \\ R_2 &\leq H(Y_2|Y_3, S) \\ R_3 &\leq H(Y_3|S) \end{aligned} \quad (32)$$

*Achievability:* By setting $U = Y_3, V = Y_2$, and $X = Y_1$ in (29), one can obtain (32).

*Converse:* Bound on $R_1$ is the same as Proposition 3.3. Now using Fano's inequality we have

$$nR_2 \leq n\epsilon_{2n} + \sum_{i=1}^{n} I(M_2; Y_{2i}|M_3, S^{i-1}, S_i, S_{i+1}^n, Y_2^{i-1})$$

$$= n\epsilon_{2n} + \sum_{i=1}^{n} H(Y_{2i}|M_3, S^{i-1}, S_{i+1}^n, Y_2^{i-1}, S_i)$$

$$\stackrel{(a)}{=} n\epsilon_{2n} + \sum_{i=1}^{n} H(Y_{2i}|M_3, S^{i-1}, S_{i+1}^n, Y_2^{i-1}, Y_{3i}, S_i)$$

$$\leq n\epsilon_{2n} + \sum_{i=1}^{n} H(Y_{2i}|Y_{3i}, S_i),$$

where (a) follows from the deterministicness of $Y_{3i}$.

$$nR_3 \leq n\epsilon_{3n} + \sum_{i=1}^{n} I(M_3; Y_{3i}|S^n, Y_3^{i-1})$$

$$= n\epsilon_{3n} + \sum_{i=1}^{n} H(Y_{3i}|S^n, Y_3^{i-1})$$

$$\leq n\epsilon_{3n} + \sum_{i=1}^{n} H(Y_{3i}|S_i).$$

∎

*Proposition 3.4:* The capacity region of deterministic 3-receiver less noisy BC with SI non-causally available at the transmitter and at the receivers $Y_1$ and $Y_2$, is the set of all rate triples $(R_1, R_2, R_3)$ such that

$$\begin{aligned} R_1 &\leq H(Y_1|Y_2, S) \\ R_2 &\leq H(Y_2|Y_3, S) \\ R_3 &\leq H(Y_3|S) \end{aligned} \quad (33)$$

*Proof:* Achievability and converse, both are immediate and therefore omitted. ∎

*Proposition 3.5:* The capacity region of the 3-receiver less noisy BC with SI non-causally available at the transmitter and at the receivers $Y_1$ and $Y_2$, with two deterministic components $Y_1$ and $Y_2$, is the set of all rate triples $(R_1, R_2, R_3)$ such that

$R_1 \leq H(Y_1|Y_2,S)$
$R_2 \leq H(Y_2|U,S)$ (34)
$R_3 \leq I(U;Y_3) - I(U;S)$

***Proof:*** The achievability is immediate. For the converse, only bounds on $R_2$ and $R_3$ might be non-trivial. Therefore,

$nR_3 = H(M_3) \stackrel{(a)}{=} H(M_3|Y_3^n) + I(M_3;Y_3^n) - I(M_3;S^n)$

$= H(M_3|Y_3^n) + \sum_{i=1}^{n} I(M_3;Y_{3i}|Y_3^{i-1}) - I(M_3;S_i|S_{i+1}^n)$

$\stackrel{(b)}{\leq} n\epsilon_{3n} + \sum_{i=1}^{n} I(M_3,Y_3^{i-1};Y_{3i}) - I(M_3,S_{i+1}^n;S_i)$

$= n\epsilon_{3n} + \sum_{i=1}^{n} I(M_3,Y_3^{i-1},S_{i+1}^n;Y_{3i}) - I(S_{i+1}^n;Y_{3i}|M_3,Y_3^{i-1})$
$\quad -I(M_3,S_{i+1}^n,Y_3^{i-1};S_i) + I(Y_3^{i-1};S_i|M_3,S_{i+1}^n)$

$\stackrel{(c)}{=} n\epsilon_{3n} + \sum_{i=1}^{n} I(M_3,Y_3^{i-1},S_{i+1}^n;Y_{3i}) - I(M_3,S_{i+1}^n,Y_3^{i-1};S_i)$

$= n\epsilon_{3n} + \sum_{i=1}^{n} I(U_i;Y_{3i}) - I(U_i;S_i)$

where $U_i \triangleq (M_3,Y_3^{i-1},S_{i+1}^n)$. (a) follows from the independence of the message and side information; (b) follows from Fano's inequality, non-negativity of mutual information, and i.i.d-ness of SI; and (c) follows from Cszizár-Körner sum identity [29].

$nR_2 \leq n\epsilon_{2n} + \sum_{i=1}^{n} I(M_2;Y_{2i}|M_3,S^{i-1},S_i,S_{i+1}^n,Y_2^{i-1})$

$\leq n\epsilon_{2n} + \sum_{i=1}^{n} I(X_{2i};Y_{2i}|M_3,S^{i-1},S_i,S_{i+1}^n,Y_2^{i-1})$

$\stackrel{(a)}{\leq} n\epsilon_{2n} + \sum_{i=1}^{n} I(X_{2i};Y_{2i}|M_3,S^{i-1},S_i,S_{i+1}^n,Y_3^{i-1})$

$= n\epsilon_{2n} + \sum_{i=1}^{n} H(Y_{2i}|M_3,S^{i-1},S_i,S_{i+1}^n,Y_3^{i-1})$

$\leq n\epsilon_{2n} + \sum_{i=1}^{n} H(Y_{2i}|M_3,S_{i+1}^n,Y_3^{i-1},S_i)$

$= n\epsilon_{2n} + \sum_{i=1}^{n} H(Y_{2i}|U_i,S_i).$ ∎

where (a) follows from Lemma 1.

***Remark 3.1:*** The capacity region of proposition 3.3 is the same as that of proposition 3.4, while the capacity region of proposition 3.2 is different from that of proposition 3.5, therefore motivating us to believe that as far as a receiver is deterministic, its full knowledge of SI does not affect the capacity region.

## V. EXAMPLES

In this section, we consider Gaussian three-receiver BC with additive Gaussian SI and see that the WDP property holds for three-receiver BCs, i.e. the capacity region of the three receiver BC with additive interference when the transmitter has full knowledge of the interference is the capacity region of three receiver BC without interference as we optimize the transmission procedure. The channel model is shown in Fig. 3. The channel input $X$ has limited power to transmit, i.e.

$$\frac{1}{n}\sum_{i=1}^{n} E(X_i^2) \leq P$$

The additive interference is assumed to be common to the three channels and is distributed i.i.d according to $N(0,Q)$, i.e. a normal random variable with zero mean and variance $Q$. Each channel has also its own additive Gaussian noise distributed normally i.i.d with zero mean and corresponding variance. Also, assume that for noise powers we have $N_1 \leq N_2 \leq N_3$. If the transmitter is fully aware of the interference, it can optimize the transmitted signal as in [22] to cancel the interference manipulation in the capacity of each discrete channel. Split the channel input to three independent parts, i.e. $X_1 \sim N(0,P_1), X_2 \sim N(0,P_2)$, and $X_3 \sim N(0,P_3)$ such that $X = X_1 + X_2 + X_3$, and $P = P_1 + P_2 + P_3$. For various receivers we have

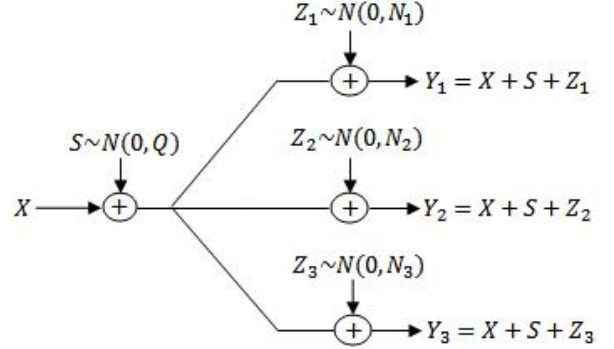

Figure 3. Gaussian three-receiver broadcast channel with additive Gaussian interference.

$Y_3 = X_3 + S + (X_1 + X_2 + Z_3)$
$Y_2 = X_2 + (S + X_3) + (X_1 + Z_2)$ (35)
$Y_1 = X_1 + (X_2 + X_3 + S) + Z_1$

We see that inputs to channels with smaller noise are themselves considered as noise for noisier channels and inputs to channels with more noise are considered as interference for channels with smaller noise due to cancellation decoding. We optimize the auxiliary random variable in each discrete channel as follows

$U_3 = X_3 + \beta_3 S_3 = X_3 + \beta_3 S$
$U_2 = X_2 + \beta_2 S_2 = X_2 + \beta_2(S + X_3)$ (36)
$U_1 = X_1 + \beta_1 S_1 = X_1 + \beta_1(X_2 + X_3 + S)$

and corresponding discrete channel capacity expressions

become

$$R_3(\beta_3) = I(U_3; Y_3) - I(U_3; S_3)$$
$$R_2(\beta_2) = I(U_2; Y_2) - I(U_2; S_2) \quad (37)$$
$$R_1(\beta_1) = I(U_1; Y_1) - I(U_1; S_1)$$

Starting from noisiest channel and maximizing the corresponding channel capacity expression to obtain optimized $\beta_k, k = 1,2,3$, we have

$$\beta_3^* = \frac{P_3}{P + N_3} \quad (38)$$

$$\beta_2^* = \frac{P_2}{P_1 + P_2 + N_2} \quad (39)$$

$$\beta_1^* = \frac{P_1}{P_1 + N_1} \quad (40)$$

and corresponding discrete channel capacities become

$$R_3(\beta_3^*) = \frac{1}{2}\log\left(1 + \frac{P_3}{P_1 + P_2 + N_3}\right) \quad (41)$$

$$R_2(\beta_2^*) = \frac{1}{2}\log\left(1 + \frac{P_2}{P_1 + N_2}\right) \quad (42)$$

$$R_1(\beta_1^*) = \frac{1}{2}\log\left(1 + \frac{P_1}{N_1}\right) \quad (43)$$

As it can be seen, there is no $Q$ in the capacity expressions.

## VI. ADDITIVE EXPONENTIAL NOISE THREE-RECEIVER BROADCAST CHANNEL WITH EXPONENTIAL DIRT

While a channel affected by numerous sources of low-power noise is modeled as a channel with Gaussian noise due to the law of large numbers, a channel afflicted by one dominant source of noise can be sometimes modeled by exponential noise especially when the dominant source of noise produces noise in nearly all of the frequency bands. This source can be the sun or cosmic rays which afflict satellite communications dominantly. Exponential noise can also model phase noises and phase interferences in phase modulation schemes. Therefore, exponential noise is of practical importance, especially in satellite-to-earth broadcast communications.

In this section, we provide simple bounds on the capacity region of the AEN-3BC-EI. The inner bound is found with the criteria that the mean value of the input signal is far more than the mean values of the noise and the interference whereas the smaller mean values are much smaller than unity and nearly close to zero. The outer bound, however, is predicated upon the equality of the mean value of the interference to the mean value of the main noise affecting the channel.

The model of the channel is depicted in Fig. 4.

Here we first provide the mentioned achievable rate for the capacity of a channel with AEN and additive exponential interference (AEI).

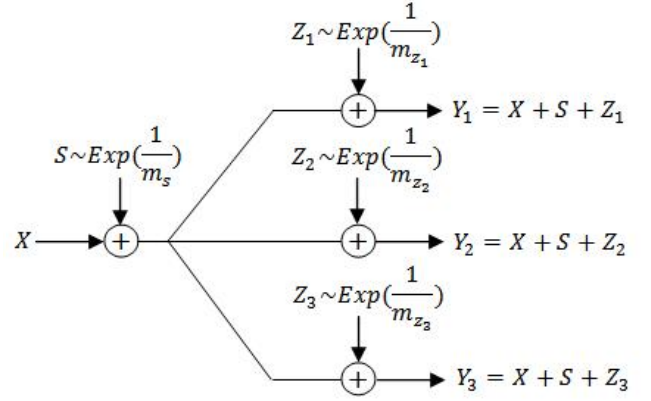

Figure 4. The additive exponential noise three-receiver broadcast channel with additive interference

First notice that according to Gelf'and-Pinsker, the capacity of a channel with side information non-causally available at the transmitter equals

$$C = \max_{p(u,x|s)} I(U; Y) - I(U; S)$$

Notice that in this case we have

$$I(U; Y) - I(U; S) = h(Y) - h(Y|U) - h(U) + h(U|S)$$

where $Y = X + S + Z$, and $h(.)$ is differential entropy with the following constraints on the input

$$X \geq 0,$$
$$EX \leq m_x.$$

Now we set $U = X + S$, then we have

$$C \geq h(X + S + Z) - h(S) - h(X + S) + h(X)$$

The output distribution which maximizes the output entropy with the constraint

$$EY \leq m_x + m_z + m_z$$

is the Exponential distribution. The input distribution that makes the output Exponential is due to [34] as follows

$$f_X(x) = \frac{m_x(m_x + m_s + m_z) + m_s m_z}{(m_x + m_s + m_z)^3} e^{-\frac{x}{m_x + m_s + m_z}} u(x)$$
$$+ \frac{(m_s + m_z)(m_x + m_s + m_z) - m_s m_z}{(m_x + m_s + m_z)^2} \delta(x)$$
$$+ \frac{m_s m_z}{m_x + m_s + m_z} \delta'(x).$$

Now with the specified distribution on $X$ and on $S$ and noticing that they are independent of each other, it is easily seen through the use of Fourier transform that $X + S$ has the following distribution

$$f_{X+S}(t) = \frac{m_x + m_s}{(m_x + m_s + m_z)^2} e^{-\frac{t}{m_x + m_s + m_z}} u(t)$$
$$+ \frac{m_z}{m_x + m_s + m_z} \delta(t).$$

The achievable rate $\Re_{m_z}$ of the above channel is

$$C \geq \ln\left(1 + \frac{m_s + m_z}{m_z}\right) + \frac{m_x(m_x + m_s + m_z) + m_s m_z}{(m_x + m_s + m_z)^2}.$$
$$\ln\left(e\left(\frac{(m_x + m_s + m_z)^3}{m_x(m_x + m_s + m_z) + m_s m_z}\right)\right)$$
$$-\frac{m_x + m_s}{m_x + m_s + m_z}\ln\left(e\left(\frac{(m_x + m_s + m_z)^2}{m_x + m_s}\right)\right)$$

provided that $m_x \gg m_s, m_x \gg m_z$, and $m_s$ and $m_z$ are both close to zero, i.e. much smaller than unity.

Therefore, we have the following Theorem

***Theorem 5:*** For $m_x \gg m_s, m_x \gg m_{z_k}, k = 1,2,3$, and $m_s$ and $m_z$ both close to zero, the set of achievable rates for the AEN-3BC-EI is given by the closure of the set of rate triples $(R_1, R_2, R_3)$ satisfying

$$R_1 \leq \Re_{m_{z_1}}$$
$$R_2 \leq \Re_{m_{z_2}}$$
$$R_3 \leq \Re_{m_{z_3}}.$$

Now we provide an outer bound on the capacity region of AEN-3BC-EI in the case that the interference has the same mean value as the main noise of the channel, i.e.

$$m_s = m_z = m$$

Now assuming that $X$ is independent of both $S$ and $Z$, and also $S$ is independent of $Z$, we see that

$$S + Z \sim Erlang(2, m)$$

that is

$$f_{S+Z}(t) = \frac{t}{m^2} e^{-\frac{t}{m}} u(t).$$

Notice that in this case

$$h(S + Z) = 2 - \psi(n)|_{n=2} + \ln(m) \tag{44}$$

where

$$\psi(z) = \frac{d}{dz}\Gamma(z), \quad \forall z \in \mathbb{R}$$

with the $\Gamma(.)$ being the Gamma function defined as

$$\Gamma(z) = \int_0^\infty t^{z-1} e^{-t} dt.$$

Therefore we have

$$\frac{d}{dz}\Gamma(z) = \frac{d}{dz}\int_0^\infty t^{z-1} e^{-t} dt = \int_0^\infty \frac{d}{dz}(t^{z-1}) e^{-t} dt$$
$$= \int_0^\infty z \ln(t) t^{z-1} e^{-t} dt = \psi(z).$$

Thus we can write

$$\psi(2) = 2 \int_0^\infty t \ln(t) e^{-t} dt = 2(1 - 0.577 \ldots) \approx 0.845568$$

Now if we put the $\psi(2)$ value in (44),

$$h(S + Z) = 1.154431 + \ln(m) \tag{45}$$

we have

$$C \leq I(X;Y) = h(Y) - h(Y|X) = h(X + S + Z) - h(S + Z)$$

The output is exponentially distributed with mean $m_x + 2m$ if

$$f_X(x) = \frac{m_x(m_x + 2m) + m^2}{(m_x + 2m)^3} e^{-\frac{x}{m_x + 2m}} u(x)$$
$$+ \frac{2m_x m + 3m^2}{(m_x + 2m)^2} \delta(x) + \frac{m^2}{m_x + 2m} \delta'(x)$$

Therefore the outer bound in this case is as follows

$$C \leq \ln(e(m_x + 2m)) - 1.154431 - \ln(m).$$

We now have the following Theorem

***Theorem 6:*** For $m_s = m_{z_k} = m, k = 1,2,3$, the set of all rate triples $(R_1, R_2, R_3)$ satisfying the following constraints, constitute an outer bound on the capacity region of AEN-3BC-EI

$$R_1 \leq \ln(e(m_x + 2m)) - 1.154431 - \ln(m)$$
$$R_2 \leq \ln(e(m_x + 2m)) - 1.154431 - \ln(m)$$
$$R_3 \leq \ln(e(m_x + 2m)) - 1.154431 - \ln(m).$$

## VII. CONCLUSION

Three-receiver broadcast channels with side information are considered. An achievable rate region for multilevel broadcast channel with side information is obtained. It is shown that the derived rate region is tight for the case where the receivers have full knowledge of side information and at least one of the receivers is a deterministic function of the input and side information. We also obtained an achievable rate region for three-receiver less noisy broadcast channel and showed that the obtained rate region is tight when side information is fully available to all the receivers. We then saw that presence of side information in deterministic receivers does not affect the capacity region. We also showed that three-receiver broadcast channels have the WDP property. Finally, we found an inner bound and an outer bound on the capacity region of additive exponential noise three receiver broadcast channel with exponential interference in special cases.